# Investigating potential energy surfaces of noncollinear molecule using variational quantum circuit

Anh Pham[1]* and Daniel Beaulieu[2]

[1]Deloitte Consulting LLP, Atlanta GA 30033

[2]Deloitte Consulting LLP, Arlington VA 22209

**Abstract**: We demonstrate the simulation of a noncollinear molecule, e.g. $H_2O$ molecule using Variational Quantum Eigensolver (VQE) with high chemical accuracy. The 2D and 3D potential energy surface (PES) were reported. Taking advantage of the potential speedup in Qiskit-runtime program, the optimal initial parameters for the variational quantum circuits were obtained after several consecutive iterations, thus resulting in accurate prediction of water's PES matching result obtained from exact diagonalization of the full Hamiltonian.

## I. Introduction and Motivation

Quantum computers hold their promise of simulating complex molecules with incredible chemical accuracies which are potentially unobtainable using the classical computing hardware. Currently, the near-term intermediate scale quantum computers (NISQs) have demonstrated their capabilities in simulating different molecular properties of simple molecules such as the ground state energy [1], potential energy surface (PES) [2], and dipole moments [3]. To predict these properties, the NISQ hardware uses hybrid quantum algorithm known as the Variational Quantum Eigensolver (VQE) to solve the Schrodinger equation without mathematical approximation as implemented in classical quantum chemistry methods like density functional theory (DFT) and Hartree-Fock (HF) approximation. The solutions obtained from the VQE method have demonstrated the chemical accuracy in comparable to advanced classical quantum chemistry methods like full configuration interactions (CI) and single and double unitary coupled cluster (UCCSD) method.

In this project, motivated by the advancement of quantum algorithms in simulating small molecules we have applied the VQE method to calculate the potential energy surface of a water molecule. $H_2O$ is a small non-collinear molecule with a bond angle of 104.5º. As a result, it has more complexity compared to the traditional diatomic molecule where there is only one degree of freedom which is the bond distance between the two atoms.

## II.  Innovation and role of Qiskit runtime

Previous study of simulating $H_2O$ on real NISQ hardware has shown the applicability of obtaining the correct electronic ground state of $H_2O$ molecule using different VQE ansatz states on ion-trapped based quantum device [4]. However, the full energy dependence of the ground state energy on the bond angle of $H_2O$ has not been studied using customized ansatz quantum circuit beyond the UCCSD-based method [5]. Taking advantage of the exponential speedup of the new *Qiskit runtime* module, we constructed a hardware efficient ansatz circuit with optimized initial rotational angles to simulate the ground state energy of $H_2O$ at different H-O-H bond angle using the VQE method with chemical accuracy matching the values obtained from exact diagonalization of the full-Hamiltonian.

## III.  Methodology and Technology stack

In this section we present the necessary steps to simulate the ground state energy of $H_2O$ at different bond angles as shown in our Jupyter notebook [6]. The initial step involves a process of constructing the Schrodinger equation in second quantization format using the classical quantum chemistry package PYSCF [7] with the STO-3G orbital basis. This minimal basis was chosen because it can be efficiently used to simulate molecule on current quantum computing hardware. In addition, in our Schrodinger equation we froze the core electrons and we also removed the three highest energy orbitals to reduce the number of qubits to speed up the run time of our simulation. Having selected the molecular orbitals for our Hamiltonian in second quantization format, we converted the Schrodinger equation to a Hamiltonian on qubit basis using the Jordan-Wigner transformation. Based on the number of available orbitals, our computation only requires 6 qubits to simulate the $H_2O$ molecule.

Subsequently, we utilize the VQE method to compute the groundstate energy of our water molecule at different initial angles using a prepared customized variational quantum circuit composed of $R_y$, $R_z$ and *CNOT* gates in the Qiskit package as the ansatz as shown in Fig. 1. The ansatz is initialized with the Hartree-Fock singlet state. Within the VQE method, the classical optimizer SPSA was chosen with a maximum iteration of 300 to ensure a convergence to the minimal energy.

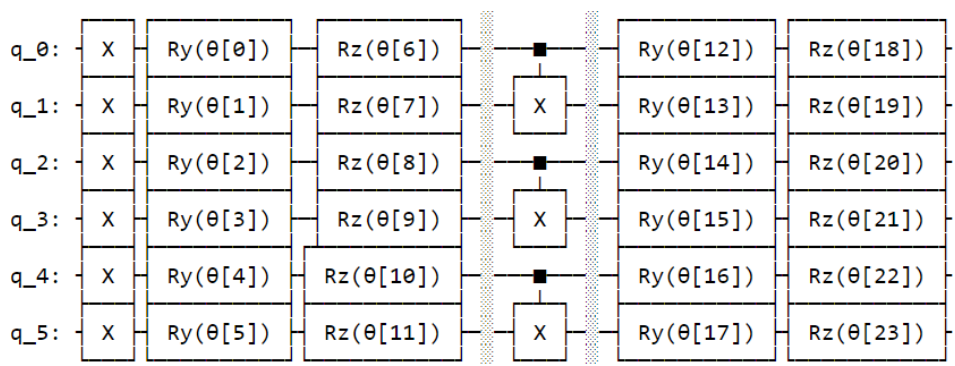

**Figure 1.** Variational quantum circuit as ansatz for water molecule simulating using VQE.

An important parameter determining the accuracy of the VQE energy is the initial starting angle in a shallow quantum circuit. To determine the optimal starting angle for our circuit, we first constructed a 3D potential energy surface using the exact diagonalization method. We then determined the optimal bond angle and the *O-H* bond length by minimizing the 3D fitted function to the PES data. With the known optimal bond length, we calculated the energy of the ground state using exact diagonalization of our qubit Hamiltonian as the benchmark energy. Next, we calculated the VQE energy at different random initial angles in our quantum circuit using the optimized molecular structure as in our exact diagonalization using the *Qiskit-runtime* program which can exponentially reduce the run time. Here we ran our simulation on the cloud *IBMQ_qasm_similator* using the *IBMQ* experience platform. By comparing the energy difference between the exact energy and the VQE energy at different runs, we obtained the optimal angles for our quantum circuit. Using these optimal angles as the initial starting points, we repeat the calculation of the potential energy surface by varying the bond angle of $H_2O$ while keeping the optimal *O-H* bond length to obtain the PES of $H_2O$ using the VQE method. Our results show that the VQE energy is consistent with the exact diagonalization when using the optimal rotational angles in the quantum circuits regardless of the molecular structures.

## IV. Results and Discussion

Water molecule is known to have angle of 104.5° and an O-H bond length of 0.945 Å based on experimental measurement. To determine the theoretical optimal angle and bond length, we constructed a 3D PES as shown in Fig. 2 by diagonalize the full Hamiltonian of $H_2O$ and fitted the data points to a 3D functions in the form of $E = \frac{1}{x^2} + \frac{1}{y^2} + \frac{1}{x^3} + \frac{1}{y^3} + \frac{1}{x^4} + \frac{1}{y^4} + \frac{1}{x^{4.5}} + \frac{1}{y^{4.5}}$ where x and y represent bond angle and the O-H bond length. By minimizing the 3D function, we obtained a minimal bond length of 108.85° and 1.00 Å respectively.

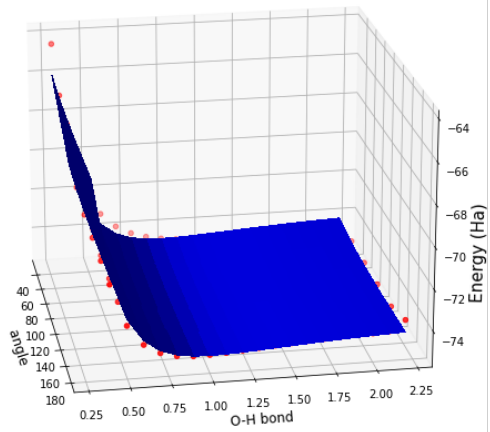

**Figure 2.** 3D Potential Energy Surface of H$_2$O molecule. The 3D data point is fitted to a function

$$E = \frac{1}{x^2} + \frac{1}{y^2} + \frac{1}{x^3} + \frac{1}{y^3} + \frac{1}{x^4} + \frac{1}{y^4} + \frac{1}{x^{4.5}} + \frac{1}{y^{4.5}}$$

Using the optimal bond length and bond angle, the grounstate energy is calculated using the VQEProgram wrapper implemented in the qiskit-runtime program. To understand the effect of the initial starting angles in the variational circuit, we ran the VQE program several times till the VQE energy matches the value from exact diagnolization of the full Hamiltonian. Within different iteration, we chose a random starting angle. As shown in Fig. 3, the VQE energy gradually approaches the Numpy result with increasing number of runs.

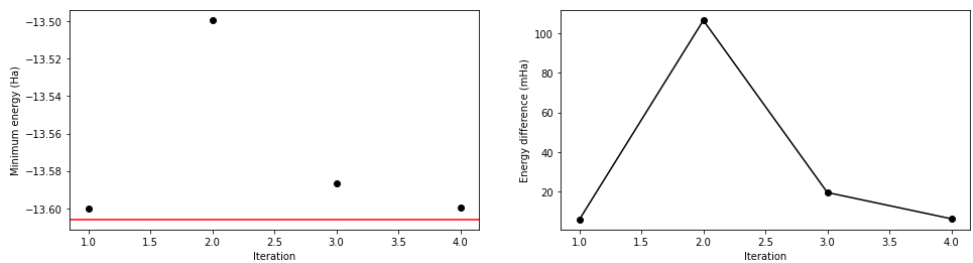

**Figure 3.** VQE energy after several runs and the comparison with the grounstate energy obtained from Numpy exact diagonalization. The red line is the reference energy from the Numpy result.

To further understand the correlation between the energy difference of the VQE result and the Numpy result, we analyzed the result from our statevectors. Interestingly, we observed that as the VQE approaches the Numpy result, the maximum probability is amplified with the increasing number of runs, and the probability difference between the most probable and second most probable outcomes increase. As a result, it implies that the VQE energy is strongly dependent on the starting angle in the variational circuit.

To construct an accurate PES for water, we utilized the optimal bond angle from the previous VQE run as the starting point and calculate the VQE groundstate energy at different bond angles. Interestingly, the VQE groundstate energy shows no dependance on the initial starting angle in the quantum circuit within different molecular structures, since the VQE energy matches the Numpy result regardless of the molecular structures using the same optimal rotational angles as the starting point in our quantum circuit.

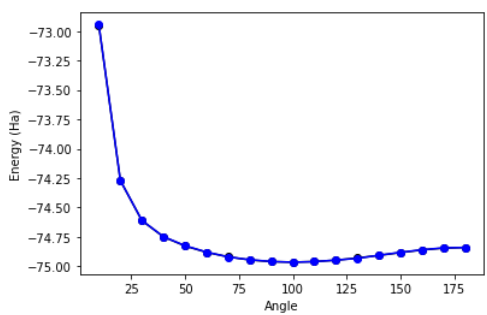

**Figure 4**. Potential energy surface of $H_2O$ at different bond angles. Here the VQE energy overlaps with the Numpy exact diagonalization.

### V. Conclusions and Future applications

In conclusion, we have demonstrated the ability of simulating the potential energy surface of a complex molecule like $H_2O$ using state of the art quantum algorithm. Using a hardware efficient quantum circuit, we were able to reproduce the chemical accuracy of our molecular simulation by first optimizing the initial starting angle for our quantum variational circuit, and utilized these parameters to construct an accurate potential energy surface for $H_2O$ using the result of exact diagonalization as our benchmark value. A future expansion of this project is to use the optimized initial starting angles to construct a 3D PES and predict the bond angle and bond lengths of $H_2O$ using *Qiskit-runtime*. In addition, another possibility is to construct a full PES using quantum machine learning with limited eigenenergies from the VQE run to accelerate the process of predicting molecular structures.